\documentclass[prd,tightenlines,nofootinbib,superscriptaddress]{revtex4}

\usepackage{amsfonts,amssymb,amsthm,bbm}

\usepackage{amsmath}

\usepackage{hyperref}

\usepackage{color,psfrag}
\usepackage[dvips]{graphicx}

\usepackage{tikz}
\usetikzlibrary{calc}
\usetikzlibrary{decorations.pathmorphing}
\usetikzlibrary{shapes.geometric}
\usetikzlibrary{arrows,decorations.markings}

\usepackage{subcaption}
\newcommand{\R}{{\mathbb R}}

\newcommand{\cL}{{\mathcal L}}
\newcommand{\cH}{{\mathcal H}}

\newcommand{\cQ}{{\mathcal Q}}

\newcommand{\cV}{{\mathcal V}}

\newcommand{\cC}{{\mathcal C}}
\newcommand{\cS}{{\mathcal S}}

\newcommand{\SL}{\mathrm{SL}}

\newcommand{\be}{\begin{equation}}
\newcommand{\ee}{\end{equation}}
\newcommand{\beq}{\begin{eqnarray}}
\newcommand{\eeq}{\end{eqnarray}}
\newcommand{\bes}{\begin{eqnarray}}
\newcommand{\ees}{\end{eqnarray}}

\renewcommand{\sl}{{\mathfrak{sl}}}

\newcommand{\f}{\frac}

\def\nn{\nonumber}
\def\pp{\partial}

\def\rd{\mathrm{d}}

\def\ka{\kappa}
\def\vphi{\varphi}
\def\eps{\epsilon}
\def\om{\omega}

\def\pip{\pi_{\phi}}

\def\ttau{\tilde{\tau}}
\def\tv{\tilde{v}}
\def\tphi{\tilde{\phi}}
\def\tN{\tilde{N}}
\def\tt{\tilde{t}}

\begin{document}

\title{The Cosmological Constant from Conformal Transformations: \vspace*{1mm}\\
M\"{o}bius Invariance and Schwarzian Action}

\author{{\bf Jibril Ben Achour}}
\affiliation{Yukawa Institute for Theoretical Physics, Kyoto University, 606-8502, Kyoto, Japan}
\author{{\bf Etera R. Livine}}\email{etera.livine@ens-lyon.fr}
\affiliation{Universit\'e de Lyon, ENS de Lyon, CNRS, Laboratoire de Physique LPENSL, 69007 Lyon, France}

\date{\today}

\begin{abstract}

The homogeneous Friedman-Lema\^\i tre-Robertson-Walker (FLRW) cosmology of a free scalar field with vanishing cosmological constant was recently shown to be invariant under the one-dimensional conformal group $\textrm{SL}(2,\mathbb{R})$ acting as M{\"o}bius transformations on the proper time. Here we generalize this analysis to arbitrary transformations of the proper time, $\tau\mapsto \tilde{\tau}=f(\tau)$, which are not to be confused with reparametrizations of the time coordinate. First, we show that the FLRW cosmology with a non-vanishing cosmological constant $\Lambda\ne 0$ is also invariant under a $\textrm{SL}(2,\mathbb{R})$ group of conformal transformations. The associated conformal Noether charges form a $\sl(2,\R)$ Lie algebra which encodes the cosmic evolution. Second, we show that a  cosmological constant can be generated from the $\Lambda=0$ case through particular conformal transformations, realizing a compactification or de-compactification of the proper time depending on the sign of $\Lambda$. Finally, we propose an extended FLRW cosmological action invariant under the full group $\textrm{Diff}({\cal S}^1)$ of conformal transformations on the proper time, by promoting the cosmological constant to a gauge field for conformal transformations or by modifying the scalar field action to a Schwarzian action.
Such a conformally-invariant cosmology leads to a renewed problem of time and to the necessity to re-think inflation in purely time-deparameterized terms.

\end{abstract}

\maketitle

\tableofcontents

\newpage

\section*{Introduction}

Symmetry is an essential concept in theoretical physics. A theory is defined as a representation of a symmetry group, which gives the constants of motion, allows for the integration of the equations of motion, constrains the quantization and the correlation functions. This role becomes even more crucial in the quest of quantum gravity, since the physical content of classical general relativity is entirely encoded in its gauge invariance under space-time diffeomorphisms and that, from this perspective, quantum gravity can be considered as the search for a (possibly anomalous) quantum representation of a regularization or extension of the group of diffeomorphisms.

In this work, we apply this logic to the homogeneous and isotropic sector of general relativity (GR). This is indeed a perfect arena to test quantum gravity methods, not only because its highly symmetric setting is simpler to handle than the generic setting of full general relativity, but also because it is the only setting (as for now) where we can reasonably hope to identify signatures of the quantum regime of gravity through the early universe dynamics.

In this context, a deeper analysis of the Friedman-Lema\^\i tre-Robertson-Walker (FLRW) cosmology of a homogeneous free scalar field coupled to general relativity with vanishing cosmological constant $\Lambda=0$ has recently identified a hidden conformal symmetry on top of the gauge invariance under standard time reparametrizations (or 1D diffeomorphisms)  \cite{BenAchour:2019ufa}.
We underline that those conformal transformations, which change the proper time, are to be distinguished from the usual time reparametrizations, which change the time coordinate while keeping the proper time invariant.
More precisely, it was shown that FLRW cosmology at $\Lambda=0$ with flat three-dimensional slices is invariant under 1D conformal group $\SL(2,\R)$ acting as M{\"o}bius transformations on the proper time,
\be
\tau\quad\longmapsto\quad\ttau=\f{\alpha\tau+\beta}{\gamma\tau+\delta}
\quad\textrm{with}\quad
\alpha,\beta,\gamma,\delta\in\R
\,.
\ee
associated to a suitable time-dependent rescaling of the scale factor.
The corresponding Noether charges form a $\sl(2,\R)$ Lie algebra, previously studied in \cite{BenAchour:2017qpb} and named the CVH algebra. This symmetry and algebraic structure allow one to map the FLRW cosmology with a free scalar field onto the conformal quantum mechanics introduced by de Alfaro, Fubini and Furlan \cite{deAlfaro:1976vlx} and to discuss it in terms of 1D conformal field theory (CFT${}_{1}$) \cite{BenAchour:2019ufa}.

The initial goal of the present work is to extend this analysis to the case of a non-vanishing cosmological constant $\Lambda\ne0$ and to understand whether de Sitter (dS) and anti de Sitter (AdS) cosmologies are also invariant under a similar one-dimensional conformal symmetry.
We show that this is indeed the case: the FLRW cosmology of a free scalar field at $\Lambda\ne0$  is invariant under M{\"o}bius transformation of the proper time composed with a map with constant Schwarzian derivative.
%
This now makes possible to discuss the quantization of FLRW cosmology with an arbitrary value of $\Lambda$, both for dS and AdS cosmology, in terms of $\SL(2,\R)$ representations and CFT${}_{1}$  as done for the $\Lambda=0$ in \cite{BenAchour:2019ufa}.

A very interesting side-product of this analysis is that those constant Schwarzian functions map the FLRW cosmology with vanishing cosmological constant onto FLRW cosmology with a non-vanishing cosmological constant. Not only this gives a new twist to the interpretation of the Schwarzian derivative as curvature, but more importantly it shows that a conformal transformation of the proper time allows to generate a cosmological constant from the free theory, at least in the homogeneous setting of FLRW cosmology.
More precisely, the action of FLRW cosmology at $\Lambda=0$ is not invariant under arbitrary conformal transformations $\tau\mapsto\ttau=f(\tau)$. Such transformations generate a potential term proportional to the Schwarzian derivative of $f$, as already noticed in \cite{BenAchour:2019ufa}. When the Schwarzian vanishes, this shows that FLRW cosmology at $\Lambda=0$ is invariant under M{\"o}bius transformations. In the case that the conformal transformation of the proper time $f(\tau)$ has a constant Schwarzian, this generates a constant potential, i.e. a cosmological constant. This seems to be in the same line of thought as the application of Korteweg-de Vries solitons to cosmology as proposed in \cite{Lidsey:2012bb}.

We push the logic of conformal invariance further and propose how to modify and extend the FLRW action principle in order to make it invariant not merely under the $\SL(2,\R)$ group of  M{\"o}bius transformations but more generally under the whole $\textrm{Diff}(\cS_{1})$ group of arbitrary conformal transformations in proper time $\tau\mapsto\ttau=f(\tau)$. We envision two different approaches.
First, in section IV, we propose to turn the cosmological constant into a cosmological field $\Lambda(\tau)$ which plays the role of a gauge field for the conformal transformations, similarly to the proposal for implementing the equivalence principle in quantum mechanics by Faraggi and Matone \cite{Faraggi_2000}. It remains to be seen if such a scenario can be derived from general relativity or conformal gravity.
Second, in section V, we propose to modify the action for the free scalar field into a Schwarzian action. The resulting conformally-invariant FLRW action can be identified with the Schwarzian boundary action for two-dimensional dilatonic gravity \`a la Jackiw-Teitelboim  \cite{Maldacena:2016upp,Engelsoy:2016xyb}. The differences lay in the physical interpretation of the model. Indeed, the AdS${}_{2}$ space-time of dilatonic gravity becomes the phase space of FLRW cosmology as explained in \cite{BenAchour:2019ufa}. Moreover, it is not yet clear how the FLRW cosmology action could be understood as a boundary theory.
Finally, in section VI, we solve explicitly the equations of motion for this new  Schwarzian action for conformally-invariant FLRW cosmology and underlines that it brings the problem of time right back at the heart of cosmology.

At the end of the day, we hope that this investigation of the conformal symmetry of FLRW cosmology can open the door to a systematic study of the quantization of FLRW cosmology in CFT terms, but also that the fully conformal invariant versions of the FLRW action could lead to an interesting dynamics for the early universe.

\section{FLRW Cosmology: Overview}


Let us consider the Friedman-Lema\^\i tre-Robertson-Walker (FLRW) cosmology of a homogeneous massless free scalar field $\phi$ coupled to the space-time geometry foliated by flat spatial slices. The 4d Lorentzian metric is parametrized by the lapse function $N(t)$ and the scale factor $a(t)$, and reads:
\be
\rd s^2=-N(t)^2\rd t^2 +a(t)^2 \delta_{ij}\rd x^i\rd x^j
\,.
\ee
The FLRW cosmological action is derived as the reduced Einstein-Hilbert action integrated on a fiducial spatial cell of volume $V_{\circ}$. Noting the cosmological constant $\Lambda$, one obtains, up to total derivatives:
\be
S[N,a,\phi]
=
V_{\circ}\int \rd t\,\bigg{[}
a^{3}\f{\phi'^{2}}{2N} -\f3{8\pi G}\f{a a'^{2}}{N}-\f{\Lambda}{8\pi G}N a^{3}
\bigg{]}
\,,
\ee
with the primes denoting the derivation with respect to the time coordinate,  $\phi'=\rd_{t}\phi$ for the scalar field and $a'=\rd_{t}a$ for the scale factor.
One can complete re-absorb the volume of the cosmological fiducial cell by introducing the volume variable $v=V_{\circ}a^{3}$:
\be
S[N,v,\phi]
=
\int \rd t\,\bigg{[}
v\f{\phi'^{2}}{2N} -\f1{8\pi G}\f{v'^{2}}{3vN}-\f{\Lambda}{8\pi G}N v
\bigg{]}
\,.
\ee
We proceed to the canonical analysis by defining the conjugate momenta:
\be
\pi_{\phi}
=
\f{\pp\cL}{\pp\phi'}
=
v\f{\phi'}N
\,,\qquad
b
=
-\f{\pp\cL}{\pp v'}
=
\f1{12\pi G}\f{v'}{Nv}
\,.
\ee
Notice the opposite convention for the volume variable.
Performing the Legendre transform, we obtain the Hamiltonian form of the action:
\be
\label{FLRWHamiltonian}
S[N,v,\phi]
=
\int \rd t\,\Big{[}
\pi_{\phi}\phi' - b v' 
-{N\cH}
\Big{]}
\qquad\textrm{with}\quad
\cH
=
\f12\bigg{[}
\f{\pi_{\phi}^{2}}{v}-\ka^{2}vb^{2}+\f{3\Lambda}{\ka^{2}}v
\bigg{]}
\,,
\ee
where we have introduced the notation $\ka=\sqrt{12\pi G}$ for the Planck length (up to a numerical factor).
The lapse $N$ is a non-dynamical variable and plays the role of a Lagrange multiplier enforcing the Hamiltonian constraint $\cH=0$.

The volume is a length cube, $[v]=L^{3}$, while its conjugate momentum is an inverse volume, $[b]=L^{-3}$. The scalar field is an inverse length, $[\phi]=L^{-1}$ and $[\pi_{\phi}]=L$. Since the cosmological constant is a curvature, $[\Lambda]=L^{-2}$, and the lapse is dimensionless, $[N]=1$, one easily checks that the Hamiltonian constraint $\cH$ is an inverse length as expected, $[\cH]=L^{-1}$. The Poisson bracket is given by the canonical structure, $\{b,v\}=\{\phi,\pip\}=1$.
The Poisson bracket with the Hamiltonian constraint gives the evolution in the proper time $\tau$, defined by absorbing the lapse factor $\rd \tau = N\rd t$ in the time coordinate:
\be
\rd_{\tau}v
=
\f1N\rd_{t}v
=
\{v,\cH\}
=
\ka^{2}vb
\,,\qquad
\rd_{\tau}\phi
=
\f\pip v
\,,\qquad
\rd_{\tau}\pip=0
\,.
\ee
In order to close the system of differential equations for the evolution, instead of computing the equation of motion for the conjugate momentum $b$, it is more convenient to introduce the observable $\cC=vb$.
This observable  generates phase space dilatation in the geometric sector, on the volume $v$ and the extrinsic curvature $b$.
Its evolution gives:
\be
\rd_{\tau}v=
\ka^{2}\cC
\,,\qquad
\rd_{\tau}\cC=\{\cC,\cH\}
=
-\cH+\f{3\Lambda}{\ka^{2}}v
\,.
\ee
Taking into account that the Hamiltonian constraint must vanish on-shell, $\cH=0=\rd_{\tau}\cH$, this gives a closed set of differential equations, which are straightforward to integrate.
This reflects that the three observables, $v$, $\cC$ and $\cH$, provided with the canonical Poisson bracket, form a closed $\sl_{2}$ Lie algebra, named the CVH algebra in \cite{BenAchour:2017qpb,BenAchour:2018jwq,BenAchour:2019ywl,BenAchour:2019ufa}. 

\medskip

Imposing that the Hamiltonian constraint vanishes, $\cH=0$, the equation of motion for the volume is a straightforward second order equation:
\be
\rd_{\tau}^{2}v=3\Lambda v
\,.
\ee
The behavior of the cosmological trajectories clearly depend on the sign of the cosmological constant:
\begin{itemize}
\item The vanishing cosmological constant case $\Lambda=0$:\\

The volume evolve linearly in proper time:
\be
\cC=\cC_{0}=\eps\f{\pip}\ka
\,,\qquad
v=\eps\ka\pip(\tau-\tau_{0})
\,,\qquad
\phi=\f\eps\ka\ln\big{[}\eps\ka^{-2}\pip(\tau-\tau_{0})\big{]}
\,,
\ee
with $v e^{\eps\ka\phi}$ remaining constant during the cosmological evolution. There is a single constant of integration, which can be interpreted as the choice of origin in proper time $\tau_{0}$.
The  arbitrary sign $\eps=\pm$ signals the contraction or expansion phase of the universe. Since the volume of the universe is meant to always remain positive, the expanding solution $\eps=+$ is valid for positive times $\tau\ge\tau_{0}$, while the contracting solution  $\eps=-$ is valid for negative times $\tau\le\tau_{0}$. The origin $\tau_{0}$ is the cosmological singularity.

\item The positive cosmological constant case $\Lambda>0$:\\

The evolution of the volume becomes exponential, with two independent solutions:
\be
\label{positivesolutionv}
v=v_{+}e^{+\om \tau}+v_{-}e^{-\om \tau}
\,,\qquad
\cC=\f\om{\ka^{2}}\Big{[}
v_{+}e^{+\om \tau}-v_{-}e^{-\om \tau}\Big{]}
\qquad\textrm{with}\quad
\om=\sqrt{3\Lambda}\,,
\ee
where $v_{\pm}$ are the two constants of integration.
So a generic solution is a superposition of contracting and expanding solutions.  The scalar field evolution can be exactly integrated:
\be
\label{positivesolutionphi}
\phi=\phi_{0}+\f\pip{\om\sqrt{v_{+}v_{-}}}\arctan\bigg{[}\sqrt{\f{v_{+}}{v_{-}}}e^{\om\tau}\bigg{]}\,.
\ee
We have regular bouncing solutions such that the volume always stays positive, 
\be
\forall \tau\in\R\,,
\qquad
v=v_{0}\cosh\om(\tau-\tau_{0})
\,,\qquad
\phi=\phi_{0}+\f{2\pip}{\om v_{0}}\arctan e^{\om(\tau-\tau_{0})}
\,,
\ee
and singular  solutions such that the volume vanishes for a finite value of the proper time,
\be
\forall \tau>\tau_{0}\,,
\qquad
v=v_{0}\sinh\om(\tau-\tau_{0})
\,,\qquad
\phi=\phi_{0}-\f{2\pip}{\om v_{0}}\text{arctanh} e^{-\om(\tau-\tau_{0})}
\,.
\ee
Both types of solutions are illustrated on the plots in fig.\ref{plot:positive}.
%
%
%
\begin{figure}[!h]
\centering
\includegraphics[width=50mm]{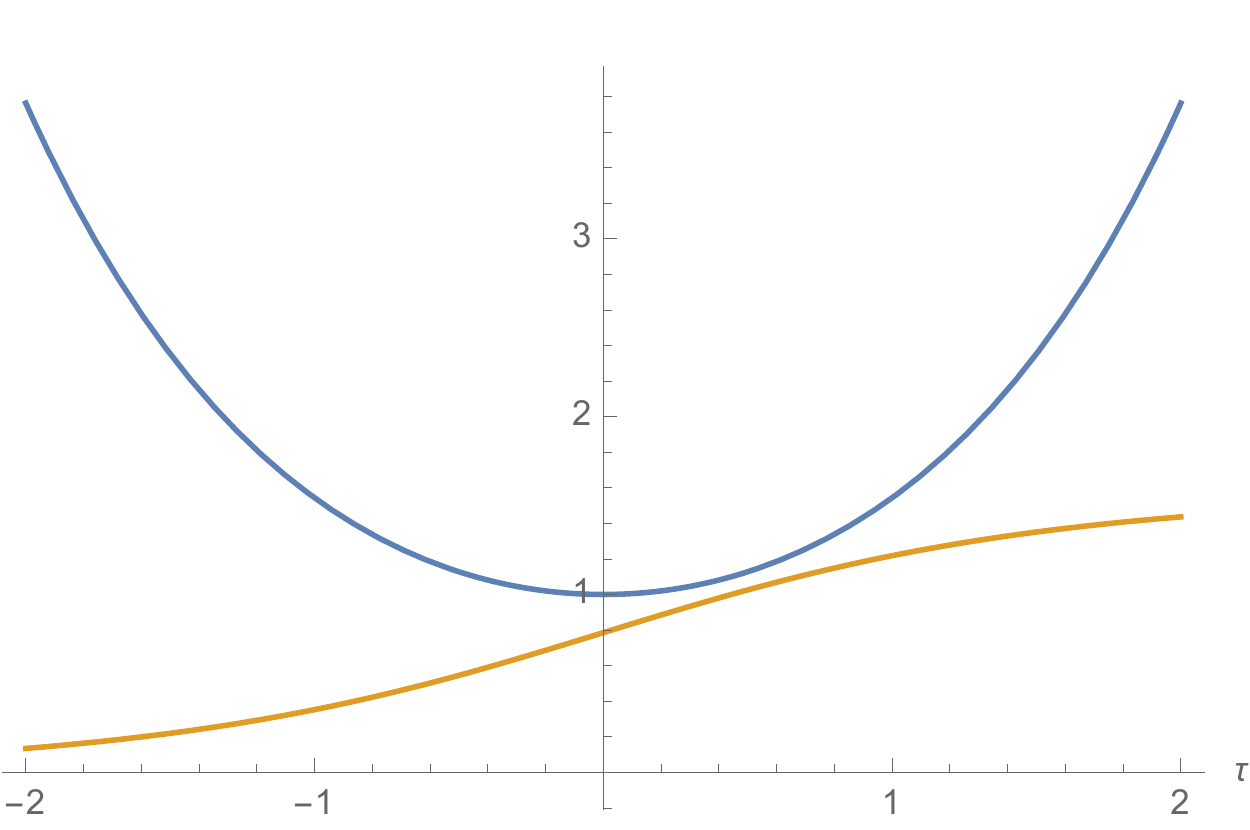}
\hspace*{20mm}
\includegraphics[width=50mm]{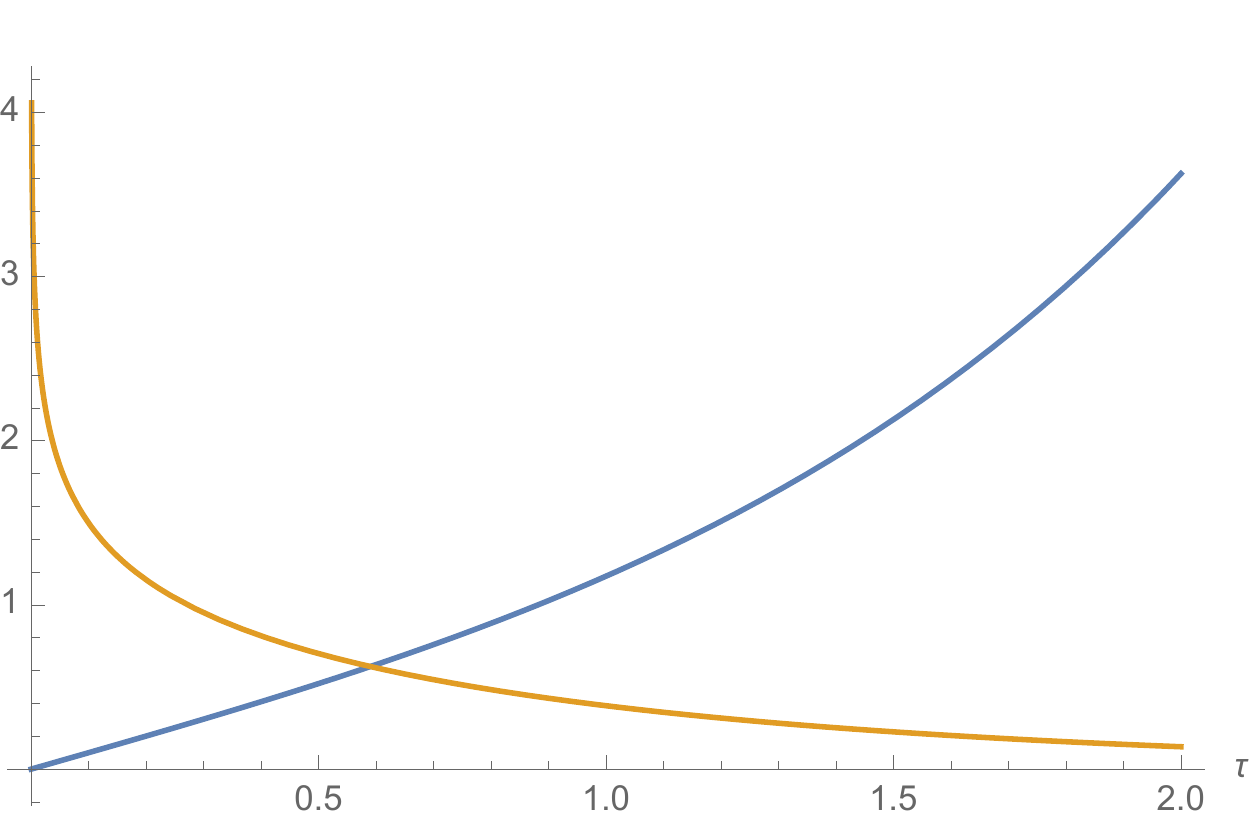}	
\caption{FLRW cosmological evolution for a positive cosmological constant $\Lambda>0$ in terms of the proper time $\tau$ with parameters $\om=1$, $\pip=\f12$, $v_{0}=1$, $\phi_{0}=\tau_{0}=0$; the volume $v$ is in blue and the scalar field $\phi$ is  in orange.
On the left hand side, a bouncing solution with $v=v_{0}\cosh\om(\tau-\tau_{0})$: the universe follows a contracting then a expanding phase while the scalar field evolves monotonically and remains bounded.
On the right hand side, a singular solution with $v=v_{0}\sinh\om(\tau-\tau_{0})$: the universe starts with an initial singularity with vanishing volume and divergent scalar field $\phi\rightarrow +\infty$, then the volume grows exponentially while the scalar field decreases toward 0.} 
\label{plot:positive}
\end{figure}

\item The negative cosmological constant case $\Lambda<0$:\\

The evolution of the volume becomes oscillatory with real frequency $\om=\sqrt{-3\Lambda}>0$:
\be
\label{negativesolutionv}
v=v_{0}\cos\om (\tau-\tau_{0})
\,,\qquad
\cC=-\f{\om v_{0}}{\ka^{2}}\,\sin\om (\tau-\tau_{0})
\,,
\ee
with the two constants of integration $v_{0}$ and $\tau_{0}$.
%
%
Since the volume $v$ is meant to stay positive, these solutions are valid only the proper time intervals $]\tau_{0}-\tfrac\pi2,\tau_{0}+\tfrac\pi2[$ of duration $\pi/\om$. The trajectory for the scalar field can also be integrated:
\be
\label{negativesolutionphi}
\phi=\phi_{0}+\f{2\pip}{\om v_{0}}
\ \text{arctanh}\bigg{[}
\tan\Big{[}
\f\om2(\tau-\tau_{0})
\Big{]}
\bigg{]}
\,,
\qquad
v\cosh\Big{[}
\f{\om v_{0}}{\pip}(\phi-\phi_{0})
\Big{]}
=v_{0}
\,,
\ee
which is again valid only during the interval $\tau\in]\tau_{0}-\tfrac\pi2,\tau_{0}+\tfrac\pi2[$ around the time origin $\tau_{0}$, with the scalar field monotonically growing from $-\infty$ to $+\infty$. As we can see on the plots on fig.\ref{plot:negative}, there are initial and final singularities.
%
\begin{figure}[!h]
\centering
\includegraphics[width=40mm]{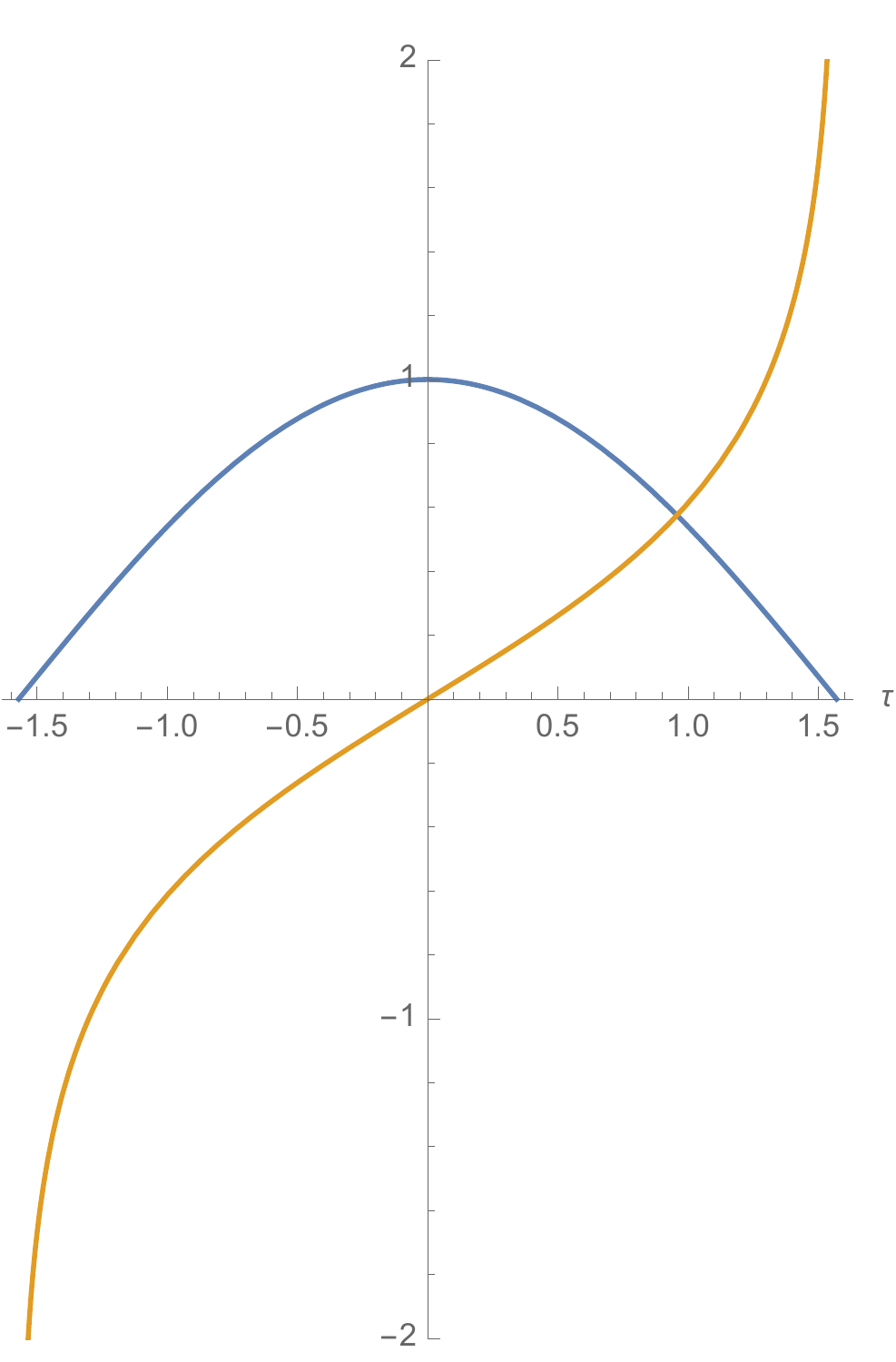}	
\caption{FLRW cosmological evolution for a negative cosmological constant $\Lambda<0$ in terms of the proper time $\tau$ with parameters $\om=1$, $\pip=\f12$, $v_{0}=1$, $\phi_{0}=\tau_{0}=0$: the dynamics starts at the initial singularity, that has vanishing volume $v\rightarrow 0$ and divergent scalar field $\phi\rightarrow-\infty$, and drives the evolution towards a final singularity with $v\rightarrow 0$ and $\phi\rightarrow+\infty$. The volume $v$ is in blue and the scalar field $\phi$ is drawn in orange.} 
\label{plot:negative}
\end{figure}

\end{itemize}

\section{(A)dS Cosmology: M\"obius invariance \& Schwarzian variation}

After this short review of FLRW cosmology with a non-vanishing cosmological constant, we would like to study its behavior under conformal transformations\footnotemark{} defined as local scale transformations in proper time, as introduced in \cite{BenAchour:2019ufa}:
\be
\label{eqn:def}
\begin{array}{rcrcl}
\tau &\mapsto & \ttau&=&f(\tau) \\
v  &\mapsto & \tv(\ttau)&=&h(\tau)\,v(\tau) \\
\phi&\mapsto & \tphi(\ttau)&=&\phi(\tau)
\end{array}
\ee
where $h(\tau)=\pp_{\tau}f$ is the Jacobian factor for  infinitesimal time variations, $\rd\ttau=h(\tau)\rd\tau$. The 3d volume $v$ has a weight $+1$ with respect to those scale transformations, while the scalar field is assumed to have a trivial transformation with vanishing weight.
\footnotetext{
It is straightforward to check that the weights of the volume $v$ and scalar field $\phi$ under conformal transformations, as given in \eqref{eqn:def} are the unique weights ensuring that the action for a vanishing cosmological constant $\Lambda=0$ is invariant under M\"obius transformations.
}

It is crucial not to confuse these conformal transformations with time reparametrizations, which change the coordinate time and the lapse without affecting the proper time. Indeed, time reparametrizations act as:
\be
\begin{array}{rcrcl}
t &\mapsto & \tt&=&\theta(t) \\
N(t) &\mapsto & \tN(\tt)&=&\dot\theta(t)^{-1}\,N(t) \\
\rd\tau  &\mapsto & \rd\ttau&=&\rd\tau \\
v&\mapsto & \tv(\tt)&=&v(t) \\
\phi&\mapsto & \tphi(\tt)&=&\phi(t)
\end{array}
\ee
and do not  change the proper time $\tau$. The conformal transformations described above are thus different from the time diffeomorphisms.
As shown in \cite{BenAchour:2019ufa}, they are conformal transformations of the 3d metric, which can be extended to the 4d space-time by acting inhomogeneously in the time and space directions.

In order to study the action of conformal transformations, it is convenient to write the FLRW action directly in proper time, by absorbing the lapse field in the time coordinate: 
\be
S[v,\phi]
=
\int \rd \tau\,\bigg{[}
v\f{\dot\phi^{2}}{2} -\f1{2\ka^{2}}\f{\dot{v}^{2}}{v}-\f{3\Lambda}{2\ka^{2}} v
\bigg{]}\,,
\ee
where the Planck length $\ka=\sqrt{12\pi G}$ plays the role of a unit of time.
The dot denotes the total variation with respect to the proper time, $\dot{v}=\rd_{\tau}v$ for the volume and $\dot{\phi}=\rd_{\tau}\phi$ for the scalar field. Let us compute the action variation  under a  conformal transformation:
\beq
\widetilde{S}[\tv,\tphi]
&=&
\int \rd \ttau\,\bigg{[}
\tv\f{(\rd_{\ttau}\tphi)^{2}}{2} -\f1{2\ka^{2}}\f{(\rd_{\ttau}v)^{2}}{\tv}-\f{3\Lambda}{2\ka^{2}} \tv
\bigg{]}
\nn\\
&=&
\int \rd \tau\,\bigg{[}
v\f{\dot\phi^{2}}{2} -\f1{2\ka^{2}}\f{\dot{v}^{2}}{v}-\f{3\Lambda}{2\ka^{2}} h^{2}v
+\f{1}{\ka^{2}}v\textrm{Sch}[f]-\f1{\ka^{2}}\f{\rd (vh^{-1}\dot{h})}{\rd\tau}
\bigg{]}\,,
\label{eq:Sf}
\eeq
where the Schwarzian derivative of $f$ appears as a volume term:
\be
\textrm{Sch}[f]
=
\f{f^{(3)}}{\dot{f}}-\f32\left(\f{\ddot{f}}{\dot{f}}\right)^{2}
=
\f{\ddot{h}}{h}-\f32\left(\f{\dot{h}}{h}\right)^{2}
=
\rd_{\tau}^{2}(\ln h) -\f12(\rd_{\tau}\ln h)^{2}
\,.
\ee
We first consider the vanishing cosmological constant action, $S_{0}$ for $\Lambda=0$. As one can see from the variation computed above, the action $S_{0}$ is invariant (up to a total derivative term) under conformal transformation with vanishing Schwarzian, i.e. M{\"o}bius transformations in proper time, as noticed earlier in \cite{BenAchour:2019ufa}:
\be
f(\tau)=M_{(\alpha,\beta,\gamma,\delta)}(\tau)=\f{\alpha \tau+\beta}{\gamma \tau+\delta}
\,,\qquad
(\alpha,\beta,\gamma,\delta)\in\R^4
\,,\qquad
\textrm{Sch}[M]=0
\,.
\ee
As shown in \cite{BenAchour:2019ufa}, this $\SL(2,\R)$ conformal invariance leads to conformal Noether charges, which have been identified as (the initial conditions) for the three observables $v$, $\cC$ and $\cH$. These Noether charges form a closed $\sl_{2}$ Lie algebra called the CVH algebra.

\medskip

In this paper, we would like to show that it is possible to extend this logic to an arbitrary non-vanishing cosmological constant, $\Lambda \ne 0$. Indeed, discarding the total derivative term, the variation of the action reads:
\be
\label{conformalmap}
\Delta_{f}S=
\int \rd \tau\,
\f{v}{\ka^{2}}\bigg{[}
\textrm{Sch}[f]
-\f{3\Lambda}{2} \left(\dot{f}^{2} -1\right)
\bigg{]}\,.
\ee
Thus the action is invariant if the conformal transformation satisfies:
\be
\textrm{Sch}[f]-\f{3\Lambda}{2} \dot{f}^{2}
=
-\f{3\Lambda}{2} 
\,.
\ee
Using the composition law for the Schwarzian derivative, $\textrm{Sch}[f_{1}\circ f_{2}]=(\dot{f}_{2})^{2}\textrm{Sch}[f_{1}]\circ f_{2}+\textrm{Sch}[f_{2}]$ (which is the cocycle equation for the Schwarzian), we can re-absorb the $\dot{f}^{2}$ term by a suitable composition.
Choosing a function $F$ with constant Schwarzian $\textrm{Sch}[F]=-{3\Lambda}/{2}$, the condition above reduces to:
\be
\label{schwarziancomp}
\textrm{Sch}[F\circ f]
=
-\f{3\Lambda}{2} = \textrm{Sch}[M\circ F]\,.
\ee
This means that if $f$ is a M{\"o}bius transformation $M$ up to conjugation (by composition) with the  function $F$ of constant Schwarzian, more explicitly $F\circ f =M\circ F$, then the FLRW action $S$ is invariant under the conformal transformation $f$. We are left with determining the existence of a map $F$ with constant Schwarzian derivative.
We will show in the next section that such maps $F(\tau)$ are given by $\tanh \f\tau2\sqrt{3\Lambda}$ or $\tan \f\tau2\sqrt{-3\Lambda}$ depending on the sign of the cosmological constant respectively for $\Lambda>0$ and $\Lambda<0$. These maps realize a compactification or de-compactification of the proper time.
This is the first key result of this short paper: the hidden symmetry of the de Sitter and Anti de Sitter cosmologies of a massless scalar field given by the 1D conformal invariance under M{\"o}bius transformations. A direct consequence of this hidden Mobius symmetry in proper time is that even if one has fixed the lapse and select a clock, this clock is not defined in an absolute way, but as an equivalence class of proper times related by $\SL(2,\mathbb{R})$ transformations, therefore relativizing the notion of proper time. As such, this finding adds an extra layer to the problem of time in standard cosmology.

For a positive cosmological constant $\Lambda >0$, this hidden $\SL(2,\R)$ symmetry of dS cosmology leads to three Noether charges:
\beq
Q_{-} & =&  \frac{1}{2} \left[ \frac{\pi^2_{\phi}}{v} - \kappa^2 v b^2 + \frac{4\Omega^2}{\kappa^2} v \right]\,, \\
Q_0 & = &\frac{\sinh{\left( 2\Omega \tau \right)}}{2\Omega} Q_{-}  + \frac{1}{\kappa^2} \; v' \; \cosh{\left( 2\Omega \tau \right)}  - \frac{2 \Omega}{\kappa^2} \; v \;  \sinh{\left( 2\Omega \tau \right)} \,, \\
Q_{+} & = &- \frac{\sinh^2{\left( \Omega \tau \right)}}{\Omega^2} Q_{-} - \frac{2}{\kappa^2} \; v' \;  \frac{\sinh{\left( 2\Omega \tau \right)} }{2\Omega}+ \frac{2}{\kappa^2} \; v \cosh{\left( 2\Omega \tau \right)}\,,
\eeq
with the frequency $\Omega=\sqrt{3\Lambda}/2$.
Using the Friedman equations, it is easy to show that these charges are indeed conserved.
These three constants of motion $Q_{-},Q_{0},Q_{+}$  respectively generate the translations, the deformed dilatation and the deformed special conformal transformations. The Noether charges associated to the AdS case are obtained by a Wick rotation $\Omega\rightarrow i\Omega$ implementing the sign switch  $\Lambda \rightarrow - \Lambda$.  Moreover, sending $\Lambda \rightarrow 0$, the charges given above reduce to the ones derived previously in \cite{BenAchour:2019ufa} for flat FLRW cosmology.

Moreover, as expected, these three conserved charges form a  $\sl(2,\mathbb{R})$ algebra given by
\begin{align}
\{ Q_{-}, Q_0\} = Q_{-} - \Omega Q_{+} \;, \qquad \{ Q_{+}, Q_{-}\} = 2 Q_{0} \;, \qquad \{ Q_0, Q_{+} \} = Q_{+}
\,.
\end{align}
These commutators can be  recast in the usual form by shifting the generator $Q_{-}$ to $Q_{-}-\Omega/2\,Q_{+}$. Notice that this off-shell structure holds at any time $\tau$ and encodes the whole cosmic evolution.
In fact, this underlying $\sl(2,\R)$ algebra of observables, uncovered in \cite{BenAchour:2019ufa}, not only reveals the 1D conformal invariance of the theory under M\"obius transformations but fully characterizes the cosmological model in terms of $\SL(2,\R)$ symmetry\footnotemark.
\footnotetext{
For instance, one can map the presently considered (A)dS cosmology of a massless scalar field onto the conformal particle action, which is the basic mechanical model with $\SL(2,\R)$ symmetry. As done in \cite{BenAchour:2019ufa}, we introduce the position variable $q=\sqrt{v}$ and its conjugate momentum $p=2b\sqrt{v}$. Working from the action principle \eqref{FLRWHamiltonian} written in its Hamiltonian form, one takes into account that the scalar field momentum $\pi_{\phi}$ is a constant of motion and thus focuses on the dynamics of the geometry sector to get the reduced action:
\be
S[N,q,p]=-\int \rd\tau\,
\bigg[p\dot q -\f12\Big[
\f{\ka^2}4p^2-\f{\pi_{\phi}^2}{q^2}-\f{3\Lambda}{\ka^2}q^2
\Big]
\bigg]
\,.
\nn
\ee
For $\Lambda=0$, we get the pure conformal particle action with an inverse square potential in $q^{-2}$. In the more general case considered here, this gives a mapping onto the Newton-Hooke extension of the conformal particle, with an extra quadratic potential term in $q^2$ \cite{Galajinsky:2013jma}.
}
Carried to the quantum level, this $\SL(2,\R)$ symmetry would guide us to quantize the (A)dS cosmology of a scalar field as a 1D conformal field theory and use conformal bootstrap techniques to get the overlap and correlation functions, as worked out for the vanishing cosmological constant case $\Lambda=0$ in \cite{BenAchour:2019ufa}.

\medskip

Moreover, pushing the analysis of the conformal maps and the Schwarzian composition as in \eqref{conformalmap} and \eqref{schwarziancomp} above, we get our second key result: it appears that one can generate an arbitrary cosmological constant out of the vanishing cosmological constant case. Indeed, as computed above, starting from the theory at $\Lambda=0$, let us do a conformal transformation and ignore the total derivative term, we get:
\be
\widetilde{S}_{f}
=
\int \rd \tau\,\bigg{[}
v\f{\dot\phi^{2}}{2} -\f1{2\ka^{2}}\f{\dot{v}^{2}}{v}
+\f{1}{\ka^{2}}v\textrm{Sch}[f]
+ (total\,\, derivative)\bigg{]}
\,.
\ee
The action is invariant under M{\"o}bius transformation, but a conformal transformation with a constant Schwarzian derivative creates an effective cosmological constant term. Thus, choosing a function $F$ with constant Schwarzian and calling  $\Lambda_{\textrm{eff}}=-\tfrac23\textrm{Sch}[F]$, this leads to:
\be
\widetilde{S}_{F}
=
\int \rd \tau\,\bigg{[}
v\f{\dot\phi^{2}}{2} -\f1{2\ka^{2}}\f{\dot{v}^{2}}{v}
-\f{3\Lambda_{\textrm{eff}}}{2\ka^{2}}v
+ (total\,\, derivative)\bigg{]}
\,,
\ee
and we have created a cosmological constant  by a mere conformal transformation  in proper time.

In short, this means that the cosmological constant amounts to a non-trivial conformal transformation.
A consequence is that  the cosmological trajectories for arbitrary $\Lambda$ should be obtained from the  cosmological trajectories at $\Lambda=0$ by this conformal map. We check this explicitly below in the next section.

\section{Cosmological Constant as a Conformal Soliton}

In the previous section, we showed that a non-vanishing cosmological constant  corresponds to a conformal transformation with constant Schwarzian derivative for the theory at $\Lambda=0$.
We refer to this as a {\it conformal soliton}, along the lines of \cite{Lidsey:2012bb,Lidsey:2013osa,Lidsey:2018byv} which proposed to use Korteweg-de Vries (KdV) solitons in cosmology.
This is an elegant way to generate a non-trivial four-dimensional curvature through conformal transformations in the homogeneous context of FLRW cosmology.
This is also an actual physical implementation of the {\it  Schwarzian derivative as curvature} (see \cite{schwarzian} for a synthetic overview of the Schwarzian derivative).

We underline that it realizes a non-trivial transformation of the proper time, very different from time reparametrizations. As we explain below, this corresponds to a compactification or de-compactification of the proper time depending on the sign of the cosmological constant. The two maps $\tau\mapsto \tanh (\Omega \tau)$ and $\tau\mapsto\tan (\Omega \tau)$  actually already appear in the framework of conformal quantum mechanics to switch between null $\sl(2,\R)$-generators and space-like or time-like generators \cite{deAlfaro:1976vlx}, and were also recently used to study the temperature of causal diamonds \cite{Arzano:2020thh}.

\subsection{$\Lambda>0$ from proper time compactification}

For a positive cosmological constant, we need a map with constant negative Schwarzian derivative. The condition $\textrm{Sch}[F]=-{3\Lambda}/{2}<0$ is solved by:
\be
F(\tau)=\tanh (\Omega \tau)
\quad\Rightarrow\quad
\textrm{Sch}[F]=-2\Omega^{2}\,,
\ee
with the proper frequency given by $\Omega=\sqrt{3\Lambda}/{2}$. We can compose this map with an arbitrary translation in  proper time (since this is a special case of a M{\"o}bius transformation, with vanishing Schwarzian) in order to off-set the origin in $\tau$.

So we start with the FLRW action with vanishing cosmological constant, perform the non-trivial conformal change of proper time $\ttau=F(\tau)$, and get the theory with positive cosmological constant:
\be
S_{0}[\tau,v,\phi]\quad\rightarrow\quad
\tilde{S}[\ttau,\tv,\tphi]=S_{0}[\ttau,\tv,\tphi]
=S_{\Lambda}[\tau,v,\phi]\,.
\ee
The proper time $\ttau$ for the theory with vanishing cosmological constant  corresponds to the proper time $\tau$ for the theory with $\Lambda>0$.
This implies that a $\Lambda=0$ trajectory in $\ttau$ should define a $\Lambda>0$ trajectory in $\tau$. Let us check this explicitly, starting with an expanding trajectory:
\be
\tv(\ttau)
=\ka\pip\ttau+\tv_{0}
=\ka\pip(\ttau-\ttau_{0})
\,,\qquad
\tphi(\ttau)=\ka^{-1}\ln\big{[}
\ka\pip(\ttau-\ttau_{0})
\big{]}
\,,
\ee
we perform the conformal transformation $F$ properly off-set by the initial proper times, $F(\tau-\tau_{0})=\ttau-\ttau_{0}$, and get:
\be
v(\tau)=\f{\tv(\ttau)}{\dot{F}(\tau-\tau_{0})}
=
\f{\ka\pip \tanh \Omega (\tau-\tau_{0})}{\Omega \cosh^{-2}\Omega(\tau-\tau_{0})}
\,,\qquad
\phi(\tau)=\tphi(\ttau)=
\ka^{-1}\ln\Big{[}
\ka\pip\tanh \Omega (\tau-\tau_{0})\Big{]}
\,.
\ee
This exactly matches the cosmological  solutions for $\Lambda >0$, with an initial singularity, given in \eqref{positivesolutionv} and \eqref{positivesolutionphi} with $2\Omega=\sqrt{3\Lambda}=\om$:
\be
v(\tau)
=
\f{\ka\pip}\om\sinh\om(\tau-\tau_{0})
\ee
%
If we play with the off-set in proper time $\tau_{0}$ and glue the conformal transformations of collapsing and expanding trajectories, we will obtain arbitrary linear combination $v=v_{+}e^{+\om \tau}+v_{-}e^{-\om \tau}$, and similarly for the scalar field, including the regular bouncing solutions.
%

Let us underline that the conformal map $F$ realizes a compactification: the proper time $\tau$ for the cosmological evolution with a positive cosmological constant $\Lambda>0$ gets compactified to the proper time $\ttau=F(\tau)=\tanh \f\om2\tau$ describing the evolution for a vanishing cosmological constant.

\subsection{$\Lambda<0$ from proper time decompactification}

For a negative cosmological constant, we need a map with constant positive Schwarzian derivative. The condition $\textrm{Sch}[F]=-{3\Lambda}/{2}>0$ is solved by:
\be
F(\tau)=\tan \f\om2 \tau
\quad\Rightarrow\quad
\textrm{Sch}[F]=+\f12\om^{2}\,,
\ee
with the proper frequency given by $\om=\sqrt{-3\Lambda}$, where we have anticipated the factor 2 rescaling of the frequency for the cosmological trajectories.

Off-set transformations $F(\tau-\tau_{0})=\ttau-\ttau_{0}$ map the $\Lambda=0$ trajectories in $\ttau$ into $\Lambda<0$ trajectories in $\tau$. Starting with an expanding trajectory $\tv(\ttau)=\ka\pip(\ttau-\ttau_{0})$,
we get:
\be
v(\tau)=\f{\tv(\ttau)}{\dot{F}(\tau-\tau_{0})}
=
\f{2\ka\pip \tan \f\om2 (\tau-\tau_{0})}{\om \cos^{-2}\f\om2(\tau-\tau_{0})}
=
\f{\ka\pip}{\om}\sin \om (\tau-\tau_{0})
\,,
\ee
and similarly for the scalar field $\phi$.
This exactly matches solution for $\Lambda <0$ given in \eqref{negativesolutionv} and \eqref{negativesolutionphi}.
Depending on the initial off-set, we indeed obtain arbitrary oscillations $v=v_{0}\cos\om \tau+w_{0}\sin\om \tau$ and similarly for the scalar field.

For a negative constant cosmological,  the conformal map $F$ now realizes a de-compactification: the proper time $|\tau|\le\tfrac\pi\om$ with finite range
for the cosmological evolution with a positive cosmological constant $\Lambda<0$ gets de-compactified to the proper time $\ttau=F(\tau)=\tan \f\om2\tau\in\R$ describing the evolution for a vanishing cosmological constant.

\section{Cosmological Potential as Conformal Gauge Field}

Up to now, we have focused on obtaining a constant volume term interpreted as a (effective) cosmological constant. Coming back to the action of conformal transformations on the original FLRW action, as derived earlier in \eqref{eq:Sf},
\beq
&&S[v,\phi]
=
\int \rd \tau\,\bigg{[}
v\f{\dot\phi^{2}}{2} -\f1{2\ka^{2}}\f{\dot{v}^{2}}{v}
-\f{3\Lambda}{2\ka^{2}} v
\bigg{]}
\nn\\
&&\quad\overset{\ttau=f(\tau)}\longmapsto\quad
\tilde{S}[\tv,\tphi]=S[\ttau,\tv,\tphi]
=
\int \rd \tau\,\bigg{[}
v\f{\dot\phi^{2}}{2} -\f1{2\ka^{2}}\f{\dot{v}^{2}}{v}
-\f{3}{2\ka^{2}}v\Big{(}
h^2\Lambda-\f23\textrm{Sch}[f]
\Big{)}
\bigg{]}\,,
\eeq
where we have discarded total derivatives,
we see that an arbitrary conformal transformation creates a potential term $\cQ(\tau)\propto h^2\Lambda-\tfrac23\textrm{Sch}[f]$. It is important to stress that this cosmological potential depends on the proper time and is not a self-interaction potential for the scalar field
\footnote{
It might be possible to exchange a time-dependent potential $\cQ(t)$ with a self-interaction potential $\cV[\phi]$ by choosing the scalar field $\phi$ as time coordinate. Using such a dynamical choice of time coordinate is consistent with the logic of using an internal clock in general relativity and would implement the ``de-parametrization'' of the theory (i.e. getting rid of the arbitrary time coordinate). Studying this possible mapping between $\cQ(t)$ and $\cV[\phi]$ is out of the scope of the present paper but is nevertheless a very interesting question.
}.
This means that we can derive cosmological trajectories with a (proper) time dependent potential from the bare cosmological trajectories (with or without cosmological constant) by a suitable  conformal transformation.

\medskip

We would like to  go further and propose a way to make the FLRW action fully invariant under conformal transformations. We can turn the cosmological constant $\Lambda$ into a cosmological field $\Lambda(\tau)$, which plays the role of a gauge field for the conformal transformation in proper time. More precisely, we assume that this new cosmological field transforms as:
\be
\begin{array}{rcrcl}
\tau &\mapsto & \ttau&=&f(\tau) \\
\Lambda(\tau)  &\mapsto &
\tilde{\Lambda}(\ttau)&=&h(\tau)^{-2}\,\Big{[}
\Lambda(\tau)+\f23\textrm{Sch}[f]
\Big{]}
\end{array}
\ee
This cosmological field transforms with a $-2$ weight plus a derivative term, similarly to a connection.
Then the extended FLRW action is invariant under all conformal transformations $\ttau=f(\tau)$. The action
\beq
 S^{ext}[v,\phi,\Lambda]
=
\int \rd \tau\,\bigg{[}
v\f{\dot\phi^{2}}{2} -\f1{2\ka^{2}}\f{\dot{v}^{2}}{v}
-\f{3\Lambda}{2\ka^{2}} v
\bigg{]}
\eeq
transforms as
\beq
& \widetilde{S}^{ext}[\tv,\tphi]= 
\int \rd \ttau\,\bigg{[}
\tilde{v}\f{(\rd_{\ttau}\tphi)^{2}}{2} -\f1{2\ka^{2}}\f{(\rd_{\ttau}v)^{2}}{\tv}-\f{3\tilde{\Lambda}}{2\ka^{2}} \tilde{v}
\bigg{]}
=
\int \rd \tau\,\bigg{[}
v\f{\dot\phi^{2}}{2} -\f1{2\ka^{2}}\f{\dot{v}^{2}}{v}
-\f{3\Lambda}{2\ka^{2}} v
-\f1{\ka^{2}}\f{\rd (vh^{-1}\dot{h})}{\rd\tau}
\bigg{]}
\,.\nn
\eeq
Up to the total derivative, turning the cosmological constant into a gauge field for the conformal transformations, or ``conformal connection'' in short, makes the FLRW action invariant under the conformal re-parametrizations of the proper time $\tau$. Let us stress that this is an extra symmetry on top of the usual time coordinate re-parametrization corresponding the one-dimensional diffeomorphisms in the time direction.

This procedure to make FLRW cosmology fully invariant under conformal transformation is along the lines of the equivalence postulate for quantum mechanics pushed forward in e.g. \cite{Faraggi_2000,Faraggi_1999,Bertoldi_2000,Matone:2002ex,Faraggi_2013,Faraggi_2015}. It would be interesting to investigate further if this extension of FLRW cosmology, 1. has relevant phenomenological implications (e.g. for the cosmological constant problem); 2. is comparable to quintessence models or other scenarii with a varying cosmological constant (e.g. \cite{Alexander:2018djy,Magueijo:2018jzg,Alexander:2019ctv,Alexander:2019wne}); 3. could be derived from a symmetry reduction of  Weyl gravity or any other conformally-invariant version of general relativity.

A key question is to classify the equivalence classes of cosmological fields $\Lambda(\tau)$ under conformal transformations, i.e. the conformal orbits, in order to study the resulting cosmological dynamics and evolution of each class.

\section{Schwarzian Action for the Scalar Field}

The extension of FLRW cosmology above turning the cosmological constant into a conformal connection implicitly implies a modification of gravity and/or the introduction of a new field.
Another way to make FLRW cosmology conformally-invariant is to modify the action for the scalar field $\phi$.
Again along the lines of \cite{Faraggi_2000}, we propose to use the Schwarzian action instead of the minimal coupling of a free and massless scalar field:
\be
\label{Sconf}
S^{conf}[v,\phi]
=
\f1{\ka^{2}}\int \rd \tau\,\bigg{[}
v\textrm{Sch}[\phi] -\f{\dot{v}^{2}}{2v}
\bigg{]}
\,,
\ee
where the cosmological constant is set to 0.
%

A straightforward calculation, equivalent to the composition law for the Schwarzian derivative, yields the behavior of the $\textrm{Sch}[\phi]$ factor under conformal transformations 
\be
\textrm{Sch}[\phi](\tau)
\quad\overset{\ttau=f(\tau)}\longmapsto\quad
\widetilde{\textrm{Sch}[\phi]}
=
\f{\rd_{\ttau}^3\tphi}{\rd_{\ttau}\tphi}
-\f32\left(
\f{\rd_{\ttau}^2\tphi}{\rd_{\ttau}\tphi}
\right)^2
=
h(\tau)^{-2}\Big{[}
\textrm{Sch}[\phi]-\textrm{Sch}[f]
\Big{]}\,,
\ee
with the Jacobian factor $h=\rd_{\tau}f$.
The term $\textrm{Sch}[f]$ compensates the one resulting  from the conformal transformation of the kinetic term of the volume and makes this new action invariant under arbitrary conformal transformation in proper time:
\be
S^{conf}[v,\phi]
\quad\overset{\ttau=f(\tau)}\longmapsto\quad
\widetilde{S}^{conf}[\tv,\tphi]=S^{conf}[v,\phi]\,,
\ee
up to a total derivative term.

It is clear that such a scalar field is not standard, it is not a mere addition of a self-interaction potential, it involves a third derivative of the field and it does not look like any usual inflationary ansatz (e.g. see \cite{Martin_2014}).
In this context, it would be very interesting to, 1. solve the conformally-invariant FLRW evolution and see how it differs from the standard FLRW cosmology in the semi-classical regime and close to the singularity; 2. check if it leads to an inflationary phase; 3. if there exists a regime where the scalar field behaves as standard scalar field with mass and potential; 4. if this Schwarzian action can be derived from a suitable dimensional reduction or boundary term of general relativity \`a la Jackiw-Teitelboim (see e.g. \cite{Mertens_2018}).

\medskip

%
We can refine our proposal for the conformally-invariant cosmology of a Schwarzian scalar field with two important remarks:
\begin{itemize}

\item One can add to the Schwarzian action for the scalar field \eqref{Sconf}  the usual kinetic term in $v\dot{\phi}^2$, which is conformally invariant on its own and does not affect the properties of the action under conformal transformation. The action acquires a new term and coupling:
\be
S^{conf}_{\beta}[v,\phi]
=
\f1{\ka^{2}}\int \rd \tau\,\bigg{[}
v\textrm{Sch}[\phi]+\beta \f12v\dot\phi^2 -\f{\dot{v}^{2}}{2v}
\bigg{]}
\,.
\ee
In fact, it was well-known that such a term can be re-absorbed directly in the Schwarzian derivative by the composition of $\phi$ with a $\tan$ or $\tanh$ map as the compactification and de-compactification used to generate a cosmological constant:
\be
\textrm{for }\beta>0
\,,\quad
\Phi=\tan \f{\sqrt{\beta}}2\phi \Rightarrow
\textrm{Sch}[\Phi]=\textrm{Sch}[\phi]+\f{\beta}2\dot\phi^2
\,,
\ee
\be
\textrm{for }\beta<0
\,,\quad
\Phi=\tanh \f{\sqrt{-\beta}}2\phi \Rightarrow
\textrm{Sch}[\Phi]=\textrm{Sch}[\phi]+\f{\beta}2\dot\phi^2
\,.
\ee
For instance, for $\beta>0$, one would use the conformally-invariant action:
\be
S^{conf}_{\beta}[v,\phi]
=
\f1{\ka^{2}}\int \rd \tau\,\bigg{[}
v\textrm{Sch}\Big{[}\tan \tfrac{\sqrt{\beta}}2\phi \Big{]} -\f{\dot{v}^{2}}{2v}
\bigg{]}
\,.
\ee

\item Although the Schwarzian action involves third derivatives of the scalar field, it is nevertheless possible to understand it as a second order action principle. Indeed the Schwarzian derivative of the scalar field is defined as:
\be
\textrm{Sch}[\phi]
=
\f{\phi^{(3)}}{\dot\phi}-\f32\left(\f{\ddot\phi}{\dot\phi}\right)^2
=
\rd^2_{\tau}\ln\dot\phi-\f12\big{(}\rd_{\tau}\ln\dot\phi\big{)}^2
\,.
\nn
\ee
Although the scalar field $\phi$ has a trivial conformal weight, the secondary field $\dot\phi$ behave as a field of weight $-1$ and the field $\psi\equiv\ln\dot\phi$ simply gets shifted by a conformal transformation:
\be
\label{confpsi}
\begin{array}{rcrcl}
\phi &\overset{\ttau=f(\tau)}\longmapsto & \tphi(\ttau)&=&\phi(\tau) \\
\rd_{\tau}\phi &\longmapsto & \rd_{\ttau}\tphi &=& h(\tau)^{-1}\rd_{\tau}\phi \\
\psi &\longmapsto & \widetilde{\psi}(\ttau)&=&\psi(\tau)-\ln h(\tau)
\end{array}
\ee
The conformally-invariant FLRW action can then be written as a usual second-order scalar field coupled to the cosmological metric\footnotemark{}:
\be
S^{conf}_{\beta}[v,\psi]
=
\f1{\ka^{2}}\int \rd \tau\,\bigg{[}
v\Big{(}
\ddot\psi-\f12{\dot\psi}^2
+\f\beta2e^{+2\psi}\Big{)}
-\f12\f{\dot{v}^{2}}{v}
\bigg{]}
\,,
\nn
\ee
with an exponential potential $\cV[\psi]\propto e^{+2\psi}$ similar to power-law potentials used in inflationary scenarios (see \cite{Martin_2014} for an overview of inflationary potentials). It would be enlightening to investigate the cosmological trajectories resulting from this action principle and understand their early universe behavior.
\footnotetext{
We nevertheless point out that the field equations taking $\phi$ or $\psi$ as primary field are slightly different. This is the difference between a third-order action and a second-order action. Taking the pure Schwarzian action as example, we underline the difference between the following two action principles:
\be
S[\psi]=\int \rd\tau\, \f12 \dot\psi^2
\qquad\ne\qquad
S[\phi,\psi]=\int \rd\tau\,\bigg{[} \f12 \dot\psi^2-\lambda(\dot\phi-e^{\psi})\bigg{]}
\,,\nn
\ee
where $\lambda$ is a Lagrange multiplier.
The equation of motion for the former action is simply the second-order equation $\ddot\psi=0$, while the equation of motion for the latter action is the third-order field equation for the Schwarzian action (or the dimensionally-reduced 2d Liouiville theory), $\ddot\psi=\lambda e^\psi$ with $\lambda(\tau)=\lambda_{0}$ constant.
While the solutions of $\ddot\psi=0$ are $\psi(\tau)=A\tau +B$, the solutions $\ddot\psi=\lambda_{0} e^\psi$ with non-vanishing $\lambda_{0}$ are $\psi(\tau)=-2\ln\tau+\psi_{0}$ with $\lambda_{0}e^{\psi_{0}}=2$. These are extra solutions of the third order theory compared to the second order theory and can be understood as ghost modes. See e.g. \cite{Maldacena:2016upp,Engelsoy:2016xyb} for details on the classical and quantum Schwarzian theory.
}

\end{itemize}

Let us compare in the next section this conformally-invariant Schwarzian cosmology with the more usual inflationary cosmology driven by a scalar field with non-trivial potential, and explore the role of the conformal symmetry.

\section{Conformally-invariant Cosmology: a Deeper Problem of Time ?}

We would like to investigate the physics of the new conformally-invariant FLRW cosmology we introduced above and especially understand the role of the invariance under conformal reparametrization of the proper time.
The theory can be defined in terms of the primary scalar $\phi$ or its descendant scalar field $\psi=\ln \dot\phi$ as:
\be
\label{confcosmo}
S^{conf}_{\beta}[v,\phi]
=
\f1{\ka^{2}}\int \rd \tau\,\bigg{[}
v\textrm{Sch}[\phi]+\beta \f12v\dot\phi^2 -\f{\dot{v}^{2}}{2v}
\bigg{]}
\quad\textrm{or}\quad
%
S^{conf}_{\beta}[v,\psi]
=
\f1{\ka^{2}}\int \rd \tau\,\bigg{[}
v\Big{(}
\ddot\psi-\f12{\dot\psi}^2
+\f\beta2e^{+2\psi}\Big{)}
-\f12\f{\dot{v}^{2}}{v}
\bigg{]}
\,.
\ee
It is intriguing that the kinetic term $\beta v\dot\phi^2$ in the primary field becomes the potential term $\beta e^{+2\psi}$ in the descendant field, and vice-versa the Schwarzian potential term $v\textrm{Sch}[\phi]$ in the primary field plays the role of the kinetic terms $v(\ddot\psi-\f12{\dot\psi}^2)$ for the descendant field. This underlines a subtle interplay between what should be distinguished as kinetic energy and what can be considered as the self-interaction potential.

In the following, we focus on the secondary formulation in terms of the descendent scalar field $\psi$, since its potential term does not involve time derivatives and therefore is more direct to compare with the standard action for a scalar field. Indeed, the FLRW cosmology of a scalar field $\vphi$ with non-trivial potential is defined by the reduced Einstein-Hilbert action:
\be
S[N,v,\vphi]
=
\f1{2\ka^{2}}\int \rd t\,\bigg{[}
\f{\ka^2}{2N}v{\vphi'^{2}} -N \ka^2 v \cV[\vphi]-\f{v'^{2}}{vN}
\bigg{]}
=
\f1{2\ka^{2}}\int \rd \tau\,\bigg{[}
\f{\ka^2}{2}v{\dot\vphi^{2}}-\ka^2v \cV[\vphi]-\f{\dot v^{2}}{v}
\bigg{]}
\,,
\ee
where $\cV[\vphi]$ is the scalar potential and the lapse field is absorbed in the proper time definition, $\tau=\int \rd t\, N$ . We have not explicitly included the cosmological constant term since it now simply corresponds to a constant shift of the potential.
The potential does not affect the definition of the conjugate momenta and only creates an additional contribution to the Hamiltonian constraint:
\be
S[N,v,\vphi]
=
\int \rd t\,\Big{[}
\pi_{\vphi}\vphi' - b v' 
-{N\cH}
\Big{]}
\qquad\textrm{with}\quad
\cH
=
-\f{\ka^{2}}{2}vb^{2}+\f{\pi_{\vphi}^{2}}{2v}+v \cV[\vphi]
\,.
\ee
The existence of the scalar potential breaks the invariance under conformal transformations of the proper time. It is not even invariant under its $\SL(2,\R)$ subgroup of M\"obius transformations in proper time. Indeed, under a conformal transformation of the proper time $\tau\mapsto\ttau=f(\tau)$ as defined earlier in \eqref{eqn:def}, the potential term in the Lagrangian is not left invariant. It transforms as $\tv\cV[\tilde{\vphi}]\,\rd\ttau =h^{2}v\cV[\vphi]\,\rd\tau $ in terms of the Jacobian $h=\pp_{\tau}\ttau$ and the $h^{2}$ factor can not be compensated by the other terms or be understood as a total derivative.

At the Hamiltonian level, this translates in the fact the CVH algebra does not close anymore, which in turn means that the equations of motion can not be simply integrated as second order differential equations in the volume $v$.
More precisely, the evolution in proper time is dictated by the Poisson brackets with the Hamiltonian constraint:
\be
\left|
\begin{array}{lclcl}
\rd_{\tau}\vphi&=&\{\vphi,\cH\}&=&\f{\pi_{\vphi}}{v}\,, \vspace*{1mm}\\
\rd_{\tau}\pi_{\vphi}&=&\{\pi_{\vphi},\cH\}&=&-v\pp_{\vphi}\cV \,,
\end{array}
\right.
\qquad
\left|
\begin{array}{lclcl}
\rd_{\tau}v&=&\{v,\cH\}&=&\ka^{2}bv=\ka^{2}\cC\,, \vspace*{1mm}\\
\rd_{\tau}\cC &=& \{bv,\cH\}&=&-\cH+2v\cV[\vphi]\,\underset{\cH=0}{\hat{=}}\,2v\cV[\vphi]
\,.
\end{array}
\right.
\ee
The non-constant potential implies non-trivial equations of motion for the scalar momentum $\pi_{\vphi}$ and the geometry dilatation generator $\cC$, leading to a tower of differential equations involving higher powers in the potential. Nevertheless, the generic feature of these models is that the volume now has a non-trivial acceleration (in proper time):
\be
\rd_{\tau}^{2}v=\ka^{2}\rd_{\tau}\cC=2\ka^{2}v\cV[\vphi]\,.
\ee
Therefore, as long as the potential remains positive, $\cV[\vphi]>0$, the scalar field drives the evolution and accelerates the growth of the volume, similarly to the effect of a positive cosmological constant $\Lambda>0$.
Nevertheless, the potential  $\cV[\vphi]$ does not remain constant during the evolution and the choice of potential intertwines the dynamics of the scalar field with the evolution of the volume.
Typically, for very small volumes $v$ close the initial singularity, the scalar field $\vphi$ would start at some non-zero value $\vphi_{0}$ giving a non-zero value of the potential, acting as a positive cosmological constant leading to an exponential growth of the volume $v$. Optimally, one would like a slow-roll dynamics with very small acceleration for the scalar field to create a plateau-like phase. This would be the inflationary regime. Then at later times, the scalar field $\vphi$ would relax towards its zero value and the volume growth would decelerate, settling in its semi-classical dynamics as during the current cosmological era. This scenario already happens for the simplest Gaussian potential\footnotemark, i.e. a simple mass term $\cV[\vphi]=\f12m^{2}\vphi^{2}$.
\footnotetext{
For the common choice of potential given by a single mass term, $\cV[\vphi]=\f12m^{2}\vphi^{2}$, we get the following cosmological equations of motion:
\be
\rd_{\tau}\vphi=\f{\pi_{\vphi}}{v}
\,,\quad
\rd_{\tau}\pi_{\vphi}=-m^{2}v\vphi
\,,\qquad
\rd_{\tau}v=\ka^{2}\cC
\,,\quad
\rd_{\tau}\cC=m^{2}v\vphi^{2}
\,,
\nn
\ee
implying a positive acceleration of the scale factor, $\rd_{\tau}^{2}v=\ka^{2}m^{2}v\phi^{2}>0$.
As long as the scalar field mass is very small compared to the Planck mass, $m \ll \ka$, if we initially start with a non-vanishing value of the scalar field, then the scalar field can be considered as almost constant $\phi\sim\phi_{0}$ around the initial time $\tau\gtrsim0$  and acts as a positive cosmological constant resulting in an exponentially growth of the volume:
\be
v\underset{\tau\rightarrow 0^{+}}\sim v_{0}e^{\ka m \phi_{0}\tau}
\,.
\nn
\ee
Then, the scalar field value slowly decreases and the system transitions out of the exponential growth regime.
At later times, the scalar field is exponentially suppressed and the volume shifts to a linear growth regime, as for a vanishing cosmological constant:
\be
\phi\underset{\tau\rightarrow +\infty}\sim \phi_{\infty}e^{-m\tau}
\,,\qquad
v\sim v_{\infty}\tau
\,.
\nn
\ee
We quickly check that the evolutions of motion are satisfied at leading order.
}

Here, in light of the conformally-invariant FLRW action principle \eqref{confcosmo}, a more relevant example, closer to the case at hand, is an exponential potential\footnotemark{} $\cV[\vphi]=A\,\exp(-{\sigma\ka \vphi})$, which leads to ``power law inflation'' \cite{Martin_2014}.
\footnotetext{
For an exponential potential $\cV[\vphi]=A\,\exp(-{\sigma\ka \vphi})$, the equations of motion read:
\be
\rd_{\tau}\vphi=\f{\pi_{\vphi}}{v}
\,,\quad
\rd_{\tau}\pi_{\vphi}=\ka\sigma A ve^{-\sigma\ka \vphi}
\,,\qquad
\rd_{\tau}v=\ka^{2}\cC
\,,\quad
\rd_{\tau}\cC=2Ave^{-\sigma\ka \vphi}
\,,
\nn
\ee
with the obvious constant of motion $\rd_{\tau}(2\pi_{\phi}-\ka\sigma \cC)=0$.
Writing $\cC-2\pi_{\phi}/\ka\sigma=B$ with $B$ constant, and using the Hamiltonian constraint $\cH=0$, this gives a second order non-linear differential equation on the volume:
\be
v\ddot v=\left(1-\f{\sigma^2}4\right)\dot v^2+\f{\ka^2\sigma^2B}2\dot v-\f{\ka^4\sigma^2}4B^2
\,.\nn
\ee
}
%
Although the conformally-invariant action principle \eqref{confcosmo} is very close to the standard FLRW action for a scalar field with exponential potential, the noticeable difference is the acceleration term $v\ddot{\psi}$, which can be written after integration by parts as a coupling term $\ddot v \psi$ between the scalar field and the scalar curvature. This point should probably be investigated further. As we explain below, this term is crucial for the conformal invariance of the theory, which leads to very different physics for the equations of motion. 

The Euler-Lagrange equations for the action $S^{conf}_{\beta}[v,\psi]$ read:
\be
\ddot\psi-\f12\dot\psi^2+\f{\ddot v}v-\f12\f{\dot v^2}{v^2}+\f\beta2 e^{2\psi}=0
\,,\qquad
\rd_{\tau}(\dot v+v\dot\psi)+\beta v e^{2\psi}=0
\,.
\ee
Using the latter equation, the former equation can also be written in terms of the quantity $(\dot v+v\dot\psi)=e^{-\psi}\rd_{\tau}(ve^{+\psi})$ giving the two equations of motion:
\be
\left|
\begin{array}{lcl}
\rd_{\tau}(\dot v+v\dot\psi)
&=&
-\beta ve^{2\psi}
\,,\vspace*{1mm}\\
(\dot v+v\dot\psi)^2
&=&
-\beta v^2e^{2\psi}
\,.
\end{array}
\right.
\ee
First, the coupling constant $\beta$ needs to be negative, $-\beta>0$.
Second, the conformal Noether charges resulting from the invariance of the action under proper time conformal reparametrizations are given by arbitrary powers of $\tau$ times the same conserved quantity:
\be
I=\f{(\dot v+v\dot\psi)^2}{v}+\beta v e^{2\psi}
\,.
\ee
We easily checked that this quantity is conserved and we further recognize that $I=0$ from the equations of motion above. In fact $I$ turns out to be the Hamiltonian (constraint) of the theory, which must vanish due to the conformal invariance.
Third, the two equations of motion are actually redundant\footnotemark, which is probably the most important point and constitutes the crucial difference with the standard FLRW cosmology of a scalar field.
\footnotetext{
At the Hamiltonian level, this translates by a constraint between the volume and scalar field. Indeed we compute the conjugate momenta and get a simple relation between them:
\be
b=-\f{\pp\cL}{\pp\dot v}=\dot\psi+\f{\dot v}v
\,,\quad
\pi_{\psi}=\f{\pp\cL}{\pp\dot \psi}=-(\dot v+v\dot\psi)
\,,\qquad
\pi_{\psi}+vb=0
\,.
\nn
\ee
This shows that the volume and scalar field are not independent variables, even before imposing the Hamiltonian constraint. 
}
More precisely, starting from $(\dot v+v\dot\psi)^2=-\beta v^2e^{2\psi}$ assuming that $\beta<0$, we take the square-root, then differentiate it to get:
\be
\begin{array}{lcl}
(\dot v+v\dot\psi)^2=-\beta v^2e^{2\psi}
&\quad\Rightarrow\quad&
(\dot v+v\dot\psi)=\pm\sqrt{-\beta} ve^{\psi}
\\
&\quad\Rightarrow\quad&
\rd_{\tau}(\dot v+v\dot\psi)=\pm\sqrt{-\beta} \rd_{\tau}(ve^{\psi})=\pm\sqrt{-\beta} e^{\psi}(\dot v+v\dot\psi)
\\
&\quad\Rightarrow\quad&
\rd_{\tau}(\dot v+v\dot\psi)=
\sqrt{-\beta}e^{\psi} (\sqrt{-\beta} ve^{\psi})
=-\beta ve^{2\psi}\,,
\end{array}
\ee
which gives the other equation of motion $\rd_{\tau}(\dot v+v\dot\psi)=-\beta ve^{2\psi}$.
This has the deep consequence that the two variables $v$ and $\psi$ are not both fully determined by the equations of motion supplemented by boundary conditions. Actually, one can keep the scalar field $\psi$ completely free and determine the evolution of the volume $v(\tau)$ in terms of the chosen profile $\psi(\tau)$.
To realize this, one solves the differential equation $(\dot v+v\dot\psi)^2=-\beta v^2e^{2\psi}$, choosing, say, the positive branch $(\dot v+v\dot\psi)=\sqrt{-\beta} ve^{\psi}$ by introducing the composite variable $W=ve^{\psi}$ and writing it as:
\be
\f{\rd W}W=\sqrt{-\beta}\,e^{\psi}\rd \tau.
\ee
Integrating this gives the evolution of the volume in terms of the chosen trajectory for the scalar field $\psi(\tau)$:
\be
v(\tau)=e^{-\psi}e^{\sqrt{-\beta}\int^\tau e^{\psi}}
\,.
\ee
As an example, one can consider a constant scalar field $\psi=\psi_{0}$, which leads to
\be
\psi(\tau)=\psi_{0}
\quad\Rightarrow\quad
v(\tau)=e^{-\psi_{0}}e^{\sqrt{-\beta} e^{\psi_{0}}\tau}
\,;
\ee
and an exponential profile $\psi(\tau)\propto e^{-\alpha \tau}$ for $\tau>0$ would lead to an evolution of the volume in terms of the exponential integral.

This indeterminacy of the equations of motion is a direct consequence of the symmetry under conformal reparametrizations in proper time. Indeed, the action is invariant under the transformations \eqref{confpsi} which we recall act as:
\be
\left|\begin{array}{rcrcl}
\tau &\longmapsto & \ttau&=&f(\tau)\\
v &\longmapsto & \tv(\ttau)&=&h(\tau)v(\tau) \\
\psi &\longmapsto & \widetilde{\psi}(\ttau)&=&\psi(\tau)-\ln h(\tau)
\end{array}
\right.
\ee
with the Jacobian $h=\pp_{\tau}f$.  Such transformations map classical trajectories onto classical trajectories. So starting from one solution to the equations of motion, one generates other solutions by acting on the original solution with arbitrary conformal transformations. The transformation law for the scalar field, $\widetilde{\psi}(\ttau)=\psi(\tau)-\ln h(\tau)$, with the simple shift in $\ln h$, means that the profile $\psi(\tau)$ is absolutely free (up to possible asymptotic boundary condition issues).
The non-trivial data of the theory is not in the specific trajectories $v(\tau)$ or $\psi(\tau)$ in terms of the proper time but in the relation between the geometric variable $v$ and the matter variable $\psi$. This is exactly what is usually referred to as ``problem of time''. Let us emphasize that this is the evolution in proper time $\tau$ and not in coordinate time $t$. The new invariance under reparametrization of the proper time, on top of the usual relativistic invariance under reparametrization of the coordinate time, leads to a second layer of problem of time. The physics of our newly proposed conformally-invariant FLRW cosmology is thus rather different from the usual treatment of the FLRW cosmology of a matter field. The proper time seems to lose its intrinsic physical meaning and one would need to understand the cosmological evolution and the inflation regime in an entirely deparametrized fashion.

\section*{Discussion}

We have proposed a new way to understand the cosmological constant, in the context of FLRW cosmology, as a conformal change in proper time. Such transformations, $\tau\mapsto\ttau=f(\tau)$, with a rescaling of the scale factor, are not to be confused with changes in coordinate time, a.k.a. time reparametrizations $t\mapsto \tt(t)$, which leave the proper time\footnotemark{} invariant and are a gauge symmetry of cosmology and more generally of general relativity.
\footnotetext{
While the time coordinate $t$ is unphysical and can be considered as pure gauge (up to time-like boundary conditions), the proper time $\tau$ is physical in the sense that it is the natural time associated to a physical process, the propagation of test particles (i.e. particles in the weak coupling regime).
}
We have shown how conformal transformations with constant Schwarzian derivative, that effectively compactify or decompactify the proper time, map the FLRW cosmology of a free scalar field at $\Lambda=0$ onto the FLRW cosmology with an arbitrary cosmological constant (given in terms of the value of the Schwarzian derivative). This means that, after conformal transformation, the cosmological dynamics of the scale factor and scalar matter, and the propagation of (test) particles, happen as if there were a non-vanishing cosmological constant.
This underlines a deep connection between the cosmological constant and the problem of time, i.e. the choice of a physical time coordinate in cosmology or more generally in general relativity.

Moreover, we recall that FLRW cosmology at $\Lambda=0$ is invariant under conformal transformations with vanishing Schwarzian derivative, i.e. invariant under the $\SL(2,\R)$ group of M{\"o}bius transformations of the proper time, as shown in \cite{BenAchour:2019ufa}. Here, we have extended this result to FLRW cosmology at $\Lambda\ne0$, showing that it also admits a $\SL(2,\R)$ symmetry, consisting in M{\"o}bius transformations conjugated by a compactification or decompactification map depending on the sign of $\Lambda$.
In short, we have showed that the dS and AdS cosmologies of a massless scalar field are also conformally invariant under $\SL(2,\R)$ transformations.
This opens the door to discussing the quantization of the FLRW cosmology of a scalar field in terms of $\SL(2,\R)$ and 1D conformal field theory whatever the sign and value of the cosmological constant $\Lambda$.
In particular, it would be interesting to investigate how the presence of the cosmological constant modifies the CFT two-point correlation function of flat FLRW quantum cosmology presented in \cite{BenAchour:2019ufa}.

To go further, there are two natural directions. On the one hand, in the context of cosmology, adding a potential for the scalar field or adding inhomogeneities would break the $\SL(2,\R)$ invariance and one would like to understand if the $\SL(2,\R)$ algebraic structure could still be helpful to quantize the theory and compute the correlation functions. On the other hand, we would like to investigate how our results extend to full general relativity, i.e. understand how conformal transformations in proper time for a space-like foliation could generate a cosmological constant, or more generally other polynomial terms in the curvature. This could potentially map some classes of $f(R)$ gravity back to Einstein gravity through such conformal transformations.

To conclude on a less speculative note, we have proposed a modified FLRW action principle, fully invariant under the whole $\textrm{Diff}(\cS_{1})$ group of conformal transformations in proper time (and not only under the $\SL(2,\R)$ subgroup of M{\"o}bius transformations). Assuming a vanishing cosmological constant, it consists in switching the free scalar field for a Schwarzian field:
\be
S[v,\phi]
=
\f1{\ka^{2}}\int \rd \tau\,\bigg{[}
v\f{\dot\phi^2}{2}-\f{\dot{v}^{2}}{2v}
\bigg{]}
\qquad\longrightarrow\qquad
S^{conf}[v,\phi]
=
\f1{\ka^{2}}\int \rd \tau\,\bigg{[}
v\textrm{Sch}[\phi] -\f{\dot{v}^{2}}{2v}
\bigg{]}
\,.
\ee
It is of course essential to check whether this new conformally-invariant version leads to a plausible dynamics of the early universe and realistic cosmology.
As we solve its equations of motion, it turns out that the evolution in proper time is not fully determined: one can consider the evolution in the matter field $\phi$ as completely free while the volume evolution $v(\tau)$ is then entirely determined by the chosen matter trajectory $\phi(\tau)$ and vice-versa, if one chooses a volume trajectory for the universe geometry, then it fully determines the matter evolution). This forces to think about the cosmological evolution and inflation in fully deparameterized in terms of the relation between geometry and matter without appealing to trajectories in time.
This renewed problem of time is the direct physical consequence of the conformal invariance and requires a deeper investigation.

Moreover, another intriguing remark is that this extended Schwarzian action looks very similar to the Schwarzian action arising on the boundary of 2D dilatonic gravity, as used in the study of a AdS${}_{2}$/CFT${}_{1}$ correspondence \cite{Maldacena:2016upp,Engelsoy:2016xyb,Mertens_2018},
\be
S_{Schw}[\varphi,u]=\int \rd \tau \,\varphi(\tau)\,\textrm{Sch}[u](\tau)
\,,
\ee
where $\varphi$ is the dilaton field and $u$ parametrizes the boundary conditions at infinity. Identifying $(v,\phi)$ to $(\varphi,u)$, we see that the difference is that FLRW cosmology retains a kinetic term for the volume of the universe, leading to an evolving background for the scalar field $\phi$.
In fact, it is this kinetic term for the volume that makes this version of FLRW cosmology invariant under arbitrary conformal transformation while the Schwarzian action $S_{Schw}$ is only invariant under the $\SL(2,\R)$ subgroup.
On the other hand, in the context of AdS${}_{2}$ gravity, the dilaton field is usually set to a constant, either by boundary conditions and equations of motion or by a time redefinition. This means that FLRW conformal cosmology that we propose might be obtained through a slightly different choice of boundary conditions for Jackiw-Teitelboim gravity.
In light of this relation, a natural outlook is to investigate this interplay between FLRW cosmology and the AdS${}_{2}$/CFT${}_{1}$ correspondence at the level of the quantum theory and correlation functions.

\section*{Acknowledgments}
The work of JBA was supported by Japan Society for the Promotion of Science Grants-in-Aid for Scientific Research No. 17H02890.


\bibliographystyle{bib-style}
\bibliography{QC}

\end{document}